# SRAM Alpha-SER estimation from Word-line Voltage Margin measurements: Design architecture and experimental results

G. Torrens, I. de Paúl, B. Alorda, S. Bota, J. Segura

*Abstract*—Experimental results from a 65nm CMOS commercial technology SRAM test chip reveal a linear correlation between a new electrical parameter –the word-line voltage margin ($V_{WLVM}$)– and the measured circuit alpha-SER. Additional experiments show that no other memory cell electrical robustness-related parameters exhibit such correlation. The technique proposed is based on correlating the $V_{WLVM}$ to the SER measured on a small number of circuit samples to determine the correlation parameters. Then, the remaining non-irradiated circuits SER is determined from electrical measurements ($V_{WLVM}$) without the need of additional radiation experiments. This method represents a significant improvement in time and cost, while simplifying the SER-determination methods since most of the circuits do not require irradiation. The technique involves a minor memory design modification that does not degrade circuit performance, while circuit area increase is negligible.

*Index Terms*—Soft Error Rate, SRAM, Single Event Upset, Stability, Accelerated testing, Alpha-particle Radiation.

## I. Introduction

Technology scaling exacerbates nanometer IC sensitivity to radiation, making the soft-error rate (SER) estimation of today circuits mandatory for many applications. In general, the most sensitive block to radiation of an IC is its embedded memory and, therefore, the SER of an IC is typically dominated by the values of its SRAM cores. This is the result of combining a relatively high SER per bit and a high bit count [1]. It has been reported that often more than 50% of the die area is used for embedded SRAMs in modern system-on-chip designs [1].

Real-time SER testing is a complicated and expensive technique performed at nominal circuit use conditions, requiring several hundreds or even thousands of devices to be tested to get a statistically significant result [2]. Moreover, the duration of a typical real-time SER experiment runs from several months to multiple years [1]-[4]. One alternative to this approach is accelerated testing where intense radiation sources, with a particle flux that is generally more than $10^6\times$ larger than the flux under nominal conditions, is applied [1][2]. Therefore, accelerated SER tests can be completed within days or even hours depending on the source flux [1][2]. The obtained data is then extrapolated to normal conditions, and various experiments (one for each radiation type) are required. This technique requires either having or contracting a radiation facility with various radiation source types and/or accelerators representing in general an expensive alternative. It has been also pointed that accelerated test results can be fouled by the experimental conditions like beam uniformity or by differences between the source and natural environment radiation spectrum [2].

Another limitation of SER estimation from circuit irradiation is the impossibility of radiating all the dies coming from the fab. SER levels obtained from a reduced sample of circuits can be extrapolated to all the circuits coming from the same batch/lot as long as all circuits behave more or less the same. However, as has been extensively shown, process parameter variations in nanometer technologies have an enormous impact on the transistor's characteristics, specially for minimum-sized devices like the ones used in SRAM memory cells [5].

While experimental data about the impact of parameter variations on memory SER is limited, results from a 90nm CMOS bulk technology show an impact of process variation on alpha-SER of 16%, while the neutron-SER is 30% [6]. It has been also reported that within die variations have a strong impact on critical charge ($Q_{crit}$) variations estimated on a 65 nm CMOS technology [7]. While the $Q_{crit}$ value does not give the absolute SER value (the SER value depends on other parameters like particle type, energy, angle of incidence, etc.), it provides an estimate of the soft error susceptibility, and can be used to further estimate the SER [7]. Recent works estimate from simulations that SER variation in digital logic blocks will scale from around 30% in 90 nm CMOS technologies up to 120% for the 28nm CMOS technology node [8]. Such SER variation is also expected to occur in memory circuits where devices are typically minimum sized and therefore highly subjected to process parameter variations. Moreover, since the SRAM memory cells behavior is based on feedback configurations, device mismatch has a crucial impact on cell stability and therefore on its robustness against ionizing radiation. Consequently, large SER inaccuracy predictions for a given circuit design fabricated on a specific technology, computed either from real time or accelerated SER experiments, will occur due to the expected impact of variation on SER levels. In this sense, the SER measurement on few samples coming from a fabrication batch or lot might be only indicative of the mean SER levels due to the impact of parameter variations. This motivates an increasing interest in modeling and describing the impact of variability on SER either considering analytical or simulation-based descriptions [7][9][10].

Given the practical limitations of running radiation experiments on all circuits coming from the fab, a new

This work has been supported by the European FEDER fund and the Spanish Ministry of Science and Innovation under project CICYT-TEC2011-25017. It has also received funding to support competitive research groups from the Balearic Government (2011-2013), financed jointly by FEDER fund. The authors are members of the Electronics Systems Group (GSE-UIB) of the University of the Balearic Islands (Palma de Mallorca, Spain).

approach capable of estimating the SER level of each individual circuit to account for the impact of parameter variations is highly desirable. In this work we investigate the correlation between experimental alpha-SER data to specific electrical measurements following a philosophy similar to the one proposed in [11] oriented to determine, from biasing currents, the SRAM cell Static Noise Margin (SNM) – a parameter related to the electrical cell stability. The technique proposed in this work infers the circuit alpha-SER from experimental data related to the ability of writing each memory cell of a given SRAM array. It requires routing a specific supply voltage line to the buffers driving each memory word-line decoder output, without implying any additional modification of neither the decoder nor memory array architectures. The design modification has no impact on the circuit performance and virtually no impact on area.

## II. TEST CHIP DESCRIPTION

We designed and fabricated an SRAM test chip in a 65nm CMOS commercial technology with two independent memory blocks of 16 kbit each, organized in 256 rows and 64 columns. Memory cells are formed by six-transistor (6T cell) where four transistors are used as two cross-coupled inverters to form a latch that stores one bit, while the other two control the latch access during write and read operations (Fig. 1-a).

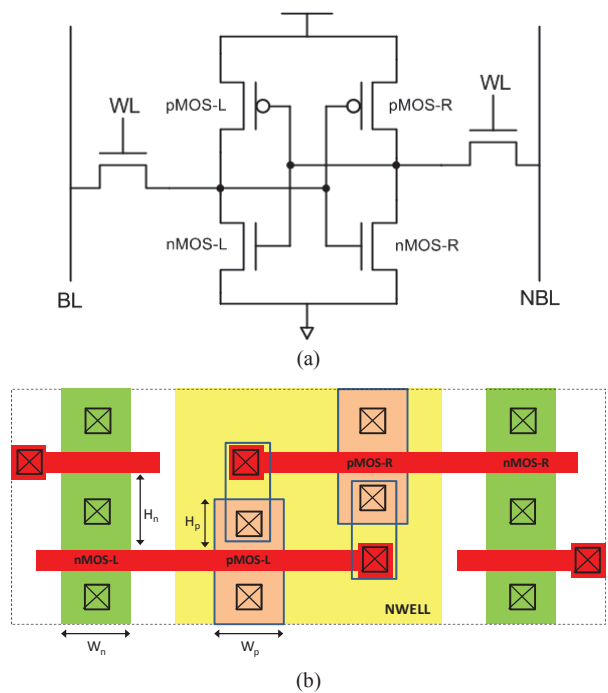

Fig. 1. Schematic and layout of a six transistors SRAM cell (6T cell).

The cells layout was implemented following the design specifications of regular-layouts [12] to minimize the impact of parameter variations. Such main layout characteristics include using straight diffusion regions and regular alignment of word-line polysilicon lines without bends as shown in Fig. 1-b.

Our design included cells with various transistor widths since from previous works it is known that they lead to memories with different robustness to soft errors from a critical charge point of view [13] or from a Soft Error Rate approach [14]. The widths of the transistors that form the cell cannot be modified arbitrarily without introducing bends in the diffusion regions and thus violating the regular-layout specifications. Transistor size changes must be performed in a way such that the diffusion regions straight shape is preserved. This goal can be achieved if all nMOS transistors widths are modified by the same factor, and if the same rule is applied to pMOS devices. If the minimum nMOS and pMOS width is taken as a reference, the two previous constraints are as follows:

$$W_n = r_n \cdot W_{\min} \qquad W_p = r_p \cdot W_{\min} \qquad (1)$$

where $W_n$ and $W_p$ are the nMOS and pMOS widths transistors respectively, $W_{\min}$ is the minimum transistor width, and $r_n$ and $r_p$ are parameters greater or equal than one. Under these two constraints, $W_n$ and $W_p$ can be modified independently while the memory cell fulfills the regular-layout specifications. These two degrees of freedom allow the designer to choose various cell designs. We included 5 cell configurations in our fabricated SRAM, represented schematically in Fig. 2.

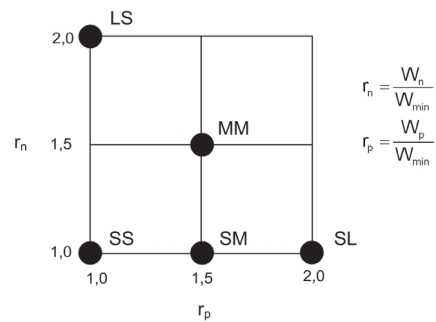

Fig. 2. 6T SRAM cell aspect ratios included in the chip.

The nomenclature for each cell indicates the relative size of the n-MOS and p-MOS transistors respectively to the minimum device width allowed by technology. Small transistors (S) are minimum sized, while medium transistors (M) width is 1.5× the minimum, and large transistors (L) width is 2× the minimum size. Therefore, for instance, an SM cell has a minimum width n-MOS transistor, and a medium sized (1.5×) p-MOS device.

The fabricated memory is organized in five 4 kb blocks, each one formed by cells of one of the five types. Fig. 3 shows a photography of the chip.

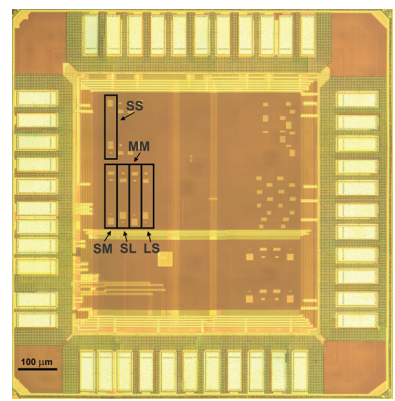

Fig. 3. Photography of the test chip showing the five memory blocks location.

Each memory block has specific inverter buffers with supply voltage lines connected to one independent supply PAD used to drive the cell word-lines providing independent word-line voltage control (Fig. 4-a). Word-line voltage modulation is a popular technique adopted in modern low voltage nanoscale SRAMs to implement write-assist (WA) techniques. WA methods are adopted to facilitate bit cell flip during the write cycle [15]. Some of these techniques allow pushing the minimum $V_{DD}$ of the memory to lower levels when this value is dictated by write failures [15].

We varied the word-line drivers supply voltage to determine the minimum word-line voltage at which each cell in the memory array could be written, $V_{WL,min}$. Table I shows these values for the considered cell types obtained from electrical simulations (typical case).

TABLE I
MINIMUM WORD-LINE VOLTAGE FOR A WRITE ACCESS
FOR THE DIFFERENT CELL TYPES

| Cell type | $V_{WL,min}$ (mV) |
|---|---|
| SS | 792 |
| SM | 853 |
| SL | 897 |
| MM | 803 |
| LS | 726 |

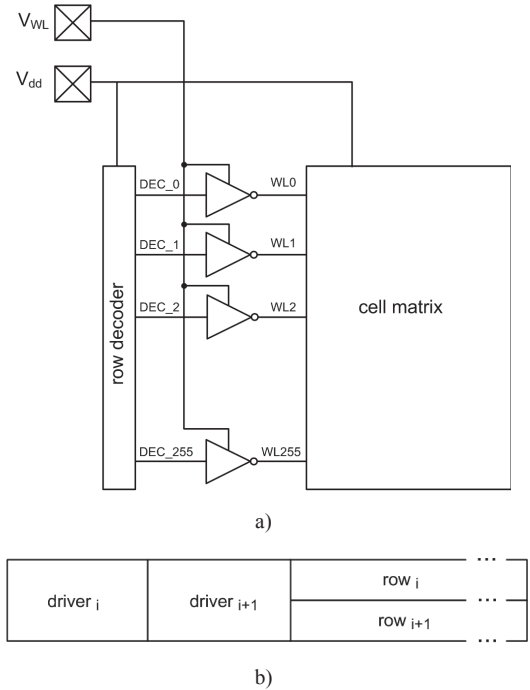

Fig. 4. a) Word-line driver schematic, and b) last decoder stage floorplan illustration.

Word-line drivers were sized according to the polysilicon word-line parasitic capacitance word-line. Note that the set of drivers used replaces the last stage of a standard word-line decoder with no word-line voltage modulation. In addition, word-line drivers were full-custom designed optimizing the area usage by sizing the drivers height to an integer number of row-cells (in our case 2 row-cells). Fig. 4-b illustrates the layout structure of a single row of word-line drivers. Each word-line driver height is equivalent to the height of two cells, and thus two drivers were included for each two rows (one for even rows and other for odd rows). The resulting layout requires routing one additional supply voltage line, and having one extra PAD. However, for all the above mentioned, the area increase of a memory implementing word-line voltage modulation is negligible when compared to a similar memory that does not implement this method. Furthermore, the technique does not impact memory performance since the whole memory activity can be executed at nominal voltage.

### III. EXPERIMENTAL RESULTS

The 65 nm test chips were mounted on a specifically designed PCB and controlled through an HP 16700 logic analysis system to drive and capture data. Three DC supply sources were used, one for the circuit IOs at 3.3V, one for the circuit logic and memory core at 1.2V and a third one for the memory word-line buffers.

#### A. Alpha SER measurements

We used an Am-241 alpha source with a 5 kBq activity providing alpha particles of 5.5 MeV. The source active area was 7 mm in diameter. The test procedure was performed following the subsequent steps:

1. Write all memory cells to a known value.
2. Read all memory cells, and compare to the written values.
3. Initiate the memory radiation.
4. Wait for a sampling time $T_s$.
5. Read the whole memory and determine the number of cells whose state changed. Go to Step 4.

Steps 4 to 5 were cycled until the experiment was finished. The overall number of SEUs ($N_{TOT}$) is computed by addition of the number of SEUs recorded at each sampling period ($N_i$), i.e.

$$N_{TOT} = \sum_{i=1}^{n} N_i \qquad (2)$$

with $n$ being the number of times that the memory is read. The overall time experiment ($t_{exp}$) is given by $t_{exp} = n\, T_s$. The SER at each sampling time period ($SER_i$) is given by $SER_i = N_i/T_s$, while the mean SER of the overall experiment is given by

$$SER = \frac{\sum_{i=1}^{n} SER_i}{n} = \frac{\sum_{i=1}^{n} N_i}{n \cdot T_S} = \frac{N_{TOT}}{t_{exp}} \qquad (3)$$

The determination of the sampling time $T_s$ is critical since it must ensure that the probability of a given cell to experience two or more flips within the sampling period is negligible, while keeping the overall read time small with respect to the overall hold time (we are interested in computing the memory hold SER). We ran an initial experiment using a one-minute $T_s$ value and determined an SER order of magnitude of 1 SEU/minute. Based on this we settled a $T_s$ value of 30 min to not increase the memory read rate. With this Ts, the mean estimated SER error due to not detected multiple flips is below 1‰.

We ran an accelerated 120 h alpha test experiment under the conditions specified. As shown in Fig. 5, the accumulated SEU count with time is linear, with the slope being the memory SER. As expected, different memory cell types in terms of transistor sizing provide different SER values.

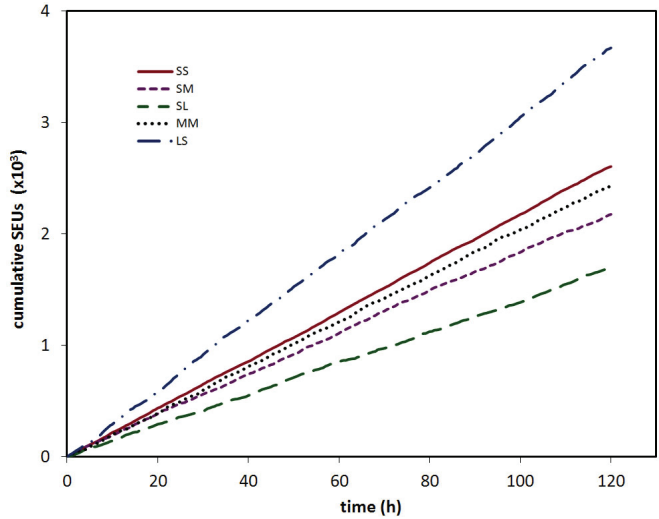

Fig. 5. Accumulated SEU count measured for each SRAM memory cell type.

SER measurements were performed for five different parts, obtaining the values reported in table II (values were normalized to the number of SEUs per second and per bit). Table II also includes the 1-sigma experimental error for each measurement arising from statistical fluctuations in the number of events. The time instants where a radiation particle induce a SEU follows a Poisson process [16], consequently their relative uncertainty is given by $1/\sqrt{N_{TOT}}$ [17], being $N_{TOT}$ the total number of SEUs.

TABLE II
SER VALUES FOR THE DIFFERENT CELL TYPES AND DIFFERENT PARTS AT NOMINAL $V_{DD}$. VALUES ARE NORMALIZED TO THE NUMBER OF SEUS PER SECOND AND PER BIT

| cell type | SER (SEUs/bit·s x10^-6) Part Number | | | | |
|---|---|---|---|---|---|
| | #1 | #2 | #3 | #4 | #5 |
| SS | 1.46±2.0% | 1.41±2.0% | 1.38±2.0% | 1.37±2.0% | 1.43±2.0% |
| SM | 1.22±2.1% | 1.16±2.2% | 1.14±2.2% | 1.14±2.2% | 1.12±2.3% |
| SL | 0.95±2.4% | 0.89±2.5% | 0.87±2.6% | 0.85±2.6% | 0.82±2.6% |
| MM | 1.38±2.0% | 1.29±2.1% | 1.32±2.1% | 1.28±2.1% | 1.25±2.1% |
| LS | 2.07±1.7% | 1.90±1.7% | 1.93±1.7% | 1.89±1.7% | 1.86±1.7% |

### B. Word-line voltage margin results

We investigated the SRAM cell stability during write for each cell type. The experiment was carried following the steps:

1. Set the word-line buffers supply voltage to its nominal value. Set counter value $n = 0$.
2. Write the whole memory array to a fixed known value with all voltages settled at nominal values, and increment the counter value $n \rightarrow n+1$.
3. Lower the word-line buffers supply voltage $V_{WL}$ by $n \cdot \Delta V_{WL}$ ($\Delta V_{WL}$ being a fixed voltage value), and write the opposite logic value with respect to Step 2.
4. Set word-line buffers supply voltage to its nominal value and read all memory cell values.
5. Compare the values read in current Step 4 to the ones read in all previous Steps 4. Determine the additional number of cells that haven't been written correctly at the current step and register this as number of write errors associated to the current $V_{WL}$ voltage.
6. Go to step 2 until all memory cells show write error.

Fig. 6 shows the percentage of write-failures associated to each word-line voltage corresponding to each memory cell type implemented in sample #1. It shows a Gaussian-like distribution, due to intra-die parameter variations, and its mean value, $V_{MEWLV}$, is defined as the mean effective word-line write voltage.

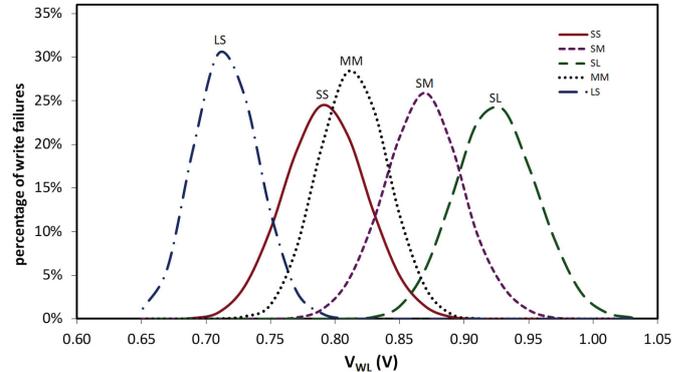

Fig. 6. Probability density function of write failures vs. the word-line voltage for each SRAM memory cell type.

The instrumental uncertainty of each individual cell word-line voltage measurement was 10 mV. However, $V_{MEWLVM}$ values were obtained by averaging 4096 cells measurements, thus getting an error associated to $V_{MEWLVM}$ below 1 mV.

Table III shows the $V_{MEWLVM}$ for all samples and their corresponding standard deviation from each of the five measured parts and for each five SRAM cell types. Part-to-part variation induces a shift in the voltage distributions, which can be quantified as a $V_{MEWLVM}$ variation between 15 and 22 mV, depending on the cell type. In addition, all distributions obtained from the same part are shifted in the same direction while the standard deviation is kept practically constant within each cell type.

TABLE III
MEAN ($V_{MEWLVM}$) AND STANDARD DEVIATION OF THE EXPERIMENTAL PROBABILITY DENSITY FUNCTIONS OF WRITE FAILURES VS. THE WORD LINE VOLTAGE.

| cell type | Part Number | | | | | | | | | |
|---|---|---|---|---|---|---|---|---|---|---|
| | #1 | | #2 | | #3 | | #4 | | #5 | |
| | μ (mv) | σ (mv) | μ (mv) | σ (mv) | μ (mv) | σ (mv) | μ (mv) | σ (mv) | μ (mv) | σ (mv) |
| SS | 791 | 44 | 795 | 42 | 806 | 42 | 803 | 43 | 814 | 44 |
| SM | 870 | 41 | 876 | 43 | 885 | 43 | 881 | 41 | 891 | 43 |
| SL | 931 | 43 | 937 | 42 | 945 | 43 | 942 | 43 | 950 | 43 |
| MM | 817 | 36 | 820 | 36 | 827 | 37 | 826 | 35 | 834 | 37 |
| LS | 715 | 32 | 717 | 33 | 723 | 32 | 723 | 33 | 730 | 35 |

The difference between the nominal supply voltage and $V_{MEWLV}$ is defined as the *Word-Line Voltage Margin* ($V_{WLVM}$). This value effectively represents the mean margin of word-line voltage lowering allowed by the memory array. This value is related to the cell strength to retain its stored value. The simulated $V_{WLVM}$ can be obtained by subtracting from $V_{DD}$ the results of Table I.

## IV. DISCUSSION

We compared the obtained $V_{WLVM}$ to the SER measured for each memory cell type. Specifically, table IV and Fig. 7 show the results corresponding to part #1, where a correlation between the measured SER and the $V_{WLVM}$ is clearly observed. Note that SER values have been normalized to the number of SEUs per second and per bit.

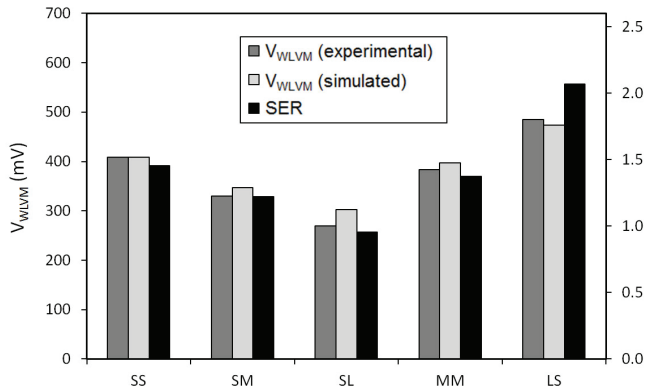

Fig. 7. Comparison between the obtained $V_{WLVM}$ and the measured SER.

TABLE IV
SER AND $V_{WLVM}$ FOR THE DIFFERENT CELL TYPES AT NOMINAL $V_{DD}$.

| Cell type | SER (SEUs/bit·s x10⁻⁶) | $V_{WLVM}$ Simulated (mV) | $V_{WLVM}$ experimental (mV) |
|---|---|---|---|
| SS | 1.46 | 408 | 409 |
| SM | 1.22 | 347 | 330 |
| SL | 0.95 | 303 | 269 |
| MM | 1.38 | 397 | 383 |
| LS | 2.07 | 474 | 485 |

The higher the word-line voltage margin, the weaker the memory cell because, as $V_{WLVM}$ increases, the memory element can be written at a smaller word-line voltage value where the influence of the bit-line voltages onto the cell internal nodes is lower. Therefore, flipping the value of a weak cell does not require the pass transistors effective drain-source resistance to be small. On the contrary, changing the stored value of a robust cell requires a strong influence of the outside bit-line values, and therefore the gate voltage of the pass transistors must be high to ensure a small as possible equivalent drain to source resistance of the pass device. This reduces significantly the word-line voltage margin, as the word line must be closer to the maximum available voltage.

To highligtht the SER–$V_{WLVM}$ correlation, Fig. 9 represents the SER results (Table II) against the experimental $V_{WLVM}$ (obtained from $V_{MEWLV}$ values of Table III) for the five parts measured at nominal $V_{DD}$. The low experimental error on the $V_{WLVM}$ measurement has been neglected. The SER error bars of Fig. 9 are obtained from the statistical uncertainties given in Table II and a second geometrical contribution related to the relative source-sample positioning that impacts the actual flux received by the circuits. This term is within ±3% for our experimental setup. The effect of this contribution is observed in Fig. 9 where the SER values of part #1 are always higher than the values of the other parts.

Experimental results reveal that the experimental SER versus $V_{WLVM}$ relationship is very close to linear and can be fitted by the equation:

$$SER = m \cdot V_{WLVM} + b \quad (4)$$

From weighted linear regression analysis for the experimental data, we obtain the fitting parameters $m = 4.32 \pm 0.20$ V⁻¹s⁻¹bit⁻¹ and $b = -0.25 \pm 0.06$ s⁻¹bit⁻¹, with $\chi^2 = 22.3$ for $\upsilon = 23$ degrees of freedom providing $\chi_\upsilon^2 = \chi^2/\upsilon = 0.97$. The resulting linear function has been represented as a solid line in Fig. 9.

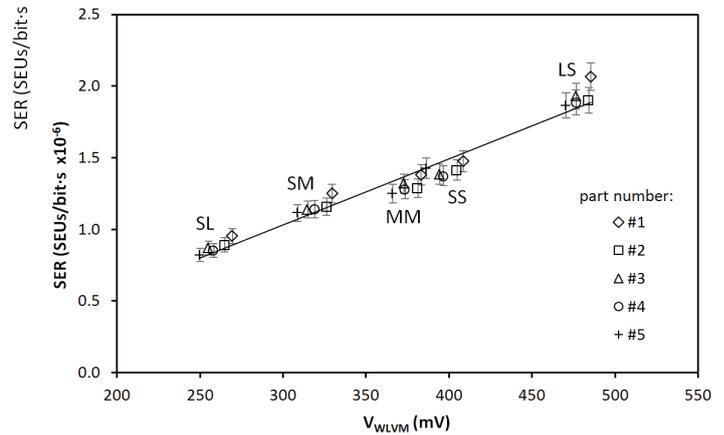

Fig. 8. Experimental SER results against $V_{WLVM}$ values. Results obtained for the five parts and for the five cell types. The $R^2$ value of the regression is 0.96.

We performed additional experimental measurements, shown in Fig. 9, to verify the linear dependence behavior for supply variation fluctuations around the nominal value. Results show that the linear dependence is maintained for ($V_{DD} \pm 10\%$).

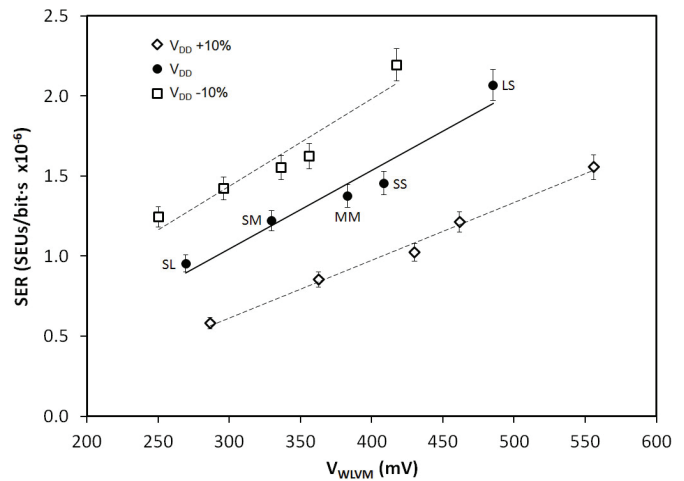

Fig. 9. Experimental SER results against $V_{WLVM}$ values, for the five cell types and for three $V_{DD}$ (nominal and ±10%).

We ran additional experiments to investigate the correlation of the measured SER to other memory robustness-related parameters. Specifically we considered:

- $V_{DD,min\ hold}$, the minimum core supply voltage at which a cell retains its stored values, and
- $V_{DD,min\ read}$, the minimum core supply voltage at which it is possible to read from a memory cell (maintaining the word-line voltage at nominal $V_{DD}$).

Both parameters have been measured following a procedure similar to the one described in Section B. However, to measure these parameters, the voltage being swept is $V_{DD}$, instead of the word-line voltage. In addition, the goal of the procedure is to determine the increment of errors for each $V_{DD}$ in hold or during read depending on the parameter being considered.

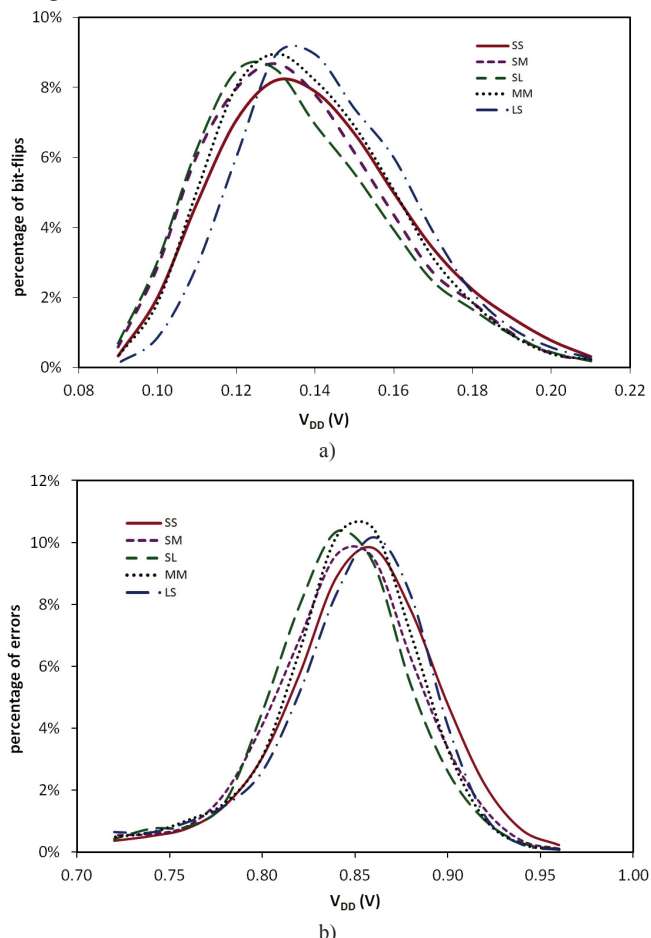

Fig. 10. Probability density function of a) bit-flips vs. supply voltage and b) read-failures vs. supply voltage for each SRAM memory cell type.

Fig. 10-a and b show the percentage of cells with a certain $V_{DD,min\ hold}$, and $V_{DD,min\ read}$ respectively. It is shown that in both cases there is a small dependence with cell aspect ratios (unlike what happens with SER and with $V_{WLVM}$). Therefore, no noticeable correlation between any of such voltage parameters and SER has been found, and the measured memory SER only exhibits a large correlation to the word-line voltage margin.

## V. CONCLUSIONS

We have experimentally established the existence of a linear relationship between the word-line voltage margin and the soft error rate due to alpha particles in CMOS SRAM memory circuits working under nominal conditions. We have also shown that $V_{WLVM}$ is a valid parameter to estimate the soft error rate of such devices, providing high savings in terms of cost. Linear fitting between SER and $V_{WLVM}$ was performed using five different cell sizes. Given that $V_{WLVM}$ can be obtained from electrical simulation, it can be used as a figure of merit at the design stage to evaluate the impact of the design parameters on the cell robustness to radiation, similarly to the use of the critical charge with the advantage that $V_{WLVM}$ can be experimentally measured easily and $Q_{crit}$ cannot. From a manufacturing point of view, the method proposed provides a quick and inexpensive technique to estimate SER once the regression coefficients between SER and $V_{WLVM}$ are determined.

The implementation cost of the method is low, as it involves minor design changes and one additional supply PAD to drive the buffer stage of row decoders, without modifying other peripheral circuitry or the memory core structure. The proposed configuration does not impact the circuit performance, while area, delay and power consumption increase are negligible.